\documentclass[12pt]{article}
\usepackage{scicite}  
\usepackage{graphicx}  
\usepackage{float}  
\usepackage{indentfirst}  
\usepackage{xcolor}  

\usepackage[
    figurename=Fig.,
    font=small,
    labelfont=bf,
    labelsep=period
]{caption}

\usepackage{amsmath}
\usepackage{amsfonts}
\usepackage{amssymb}
\usepackage{upgreek}
\usepackage{braket}

\usepackage{newtxtext,newtxmath}

\newenvironment{sciabstract}{%
\begin{quote} \bf}
{\end{quote}}

\title{Shaping terahertz waves using anisotropic shear modes \\ in a van der Waals mineral}
\author{Nicolas M. Kawahala$^1$, Daniel A. Matos$^1$, Raphaela de Oliveira$^2$,\\
Raphael Longuinhos$^3$, Jenaina Ribeiro-Soares$^3$,\\
Ingrid D. Barcelos$^2$, Felix G. G. Hernandez$^{1*}$\\
\\
\normalsize{$^1$Instituto de Física, Universidade de São Paulo, São Paulo, SP 05508-090, Brazil.}\\
\normalsize{$^2$Brazilian Synchrotron Light Laboratory, Brazilian Center for Research in}\\
\normalsize{Energy and Materials, Campinas, SP 13083-100, Brazil.}\\
\normalsize{$^3$Departamento de Física, Universidade Federal de Lavras, Lavras,}\\
\normalsize{MG 37200-900, Brazil.}\\
\\
\normalsize{*Corresponding author. Email: felixggh@if.usp.br}
}
\date{}

\begin{document}

\maketitle

\begin{sciabstract} 
    Naturally occurring van der Waals (vdW) materials are currently attracting significant interest due to their potential as low-cost sources of two-dimensional materials. Valuable information on vdW materials' interlayer interactions is present in their low-frequency rigid-layer phonon spectra, which are experimentally accessible by terahertz spectroscopy techniques. In this work, we have used polarization-sensitive terahertz time-domain spectroscopy to investigate a bulk sample of the naturally abundant, large bandgap vdW mineral clinochlore. We observed a strong and sharp anisotropic resonance in the complex refractive index spectrum near 1.13~THz, consistent with our density functional theory predictions for shear modes. Polarimetry analysis revealed that the shear phonon anisotropy reshapes the polarization state of transmitted THz waves, inducing Faraday rotation and ellipticity. Furthermore, we used Jones formalism to discuss clinochlore phononic symmetries and describe our observations in terms of its eigenstates of polarization. These results highlight the potential of exploring vdW minerals as central building blocks for vdW heterostructures with compelling technological applications.
\end{sciabstract}

\section*{INTRODUCTION}

Since the discovery of graphene \cite{Novoselov_2004}, great attention has been attracted to the research on naturally occurring van der Waals (vdW) materials \cite{Frisenda_2020}. These layered minerals are generally characterized by strongly bonded intralayer structures that are held together by weak interlayer vdW interactions \cite{Ajayan_2016}. As exfoliation \cite{Islam_2022} of the bulk crystal can be used to isolate a few of its layers, down to a single one, vdW minerals are considered a low-cost source of high-quality two-dimensional (2D) materials, where appealing quantum phenomena can emerge from low-dimensionality. Exploring 2D materials with complementary properties is crucial for the development of a variety of electronic, optoelectronic, and mechanical nanodevices that leverage different stacks of vdW heterostructures \cite{Geim_2013,Liu_2016,Castellanos-Gomez_2022}. Therefore, physically characterizing vdW minerals, particularly understanding their interlayer interactions, is essential for elucidating their properties, including optical anisotropy.

Among the family of broadly available natural vdW materials, one can highlight the mineral clinochlore, one of the most abundant phyllosilicates in nature \cite{Oliveira_2022,Oliveira_2023}. With a chemical formula Mg$_5$Al(AlSi$_3$)O$_{10}$(OH)$_8$ and a unit cell composed of an alternating stacking of brucite- and talc-like layers \cite{Spinnler_1984,Ulian_2020}, clinochlore is a large bandgap insulator that can be mechanically exfoliated down to a few layers. These properties make it a promising air-stable and optically active component for future optoelectronic devices. Although a high throughput investigation has already covered the optical characterization of clinochlore across a broad range of energies \cite{Oliveira_2022}, there is still a need for exploring its low-frequency spectrum in the terahertz (THz) band, down below 3~THz (100~cm$^{-1}$ in wavenumber). In layered crystals with more than one layer in their unit cell, this frequency range houses zone-center optical phonons associated with the relative movement of adjacent layers as rigid units \cite{Zallen_1974,longuinhos2016ultra-weak,longuinhos2024thickness}. On the contrary, in layered crystals with only one layer in their unit cell, the wavevector of the rigid-layer lattice vibrations are at the zone-edges \cite{longuinhos2020raman,longuinhos2021raman,longuinhos2021mechanical,longuinhos2023raman}. The rigid-layer modes typically appear in between 0.45~THz and 1.5~THz (15~cm$^{-1}$ and 50~cm$^{-1}$) \cite{Gasanly_1982} and are predominantly determined by the interlayer interactions \cite{Nakashima_1979}. Experimentally accessing these modes can provide insightful information on the interlayer interacting forces involved.  

THz spectroscopy techniques are well-suited for investigating lattice dynamics in the low-energy range, especially for identifying infrared (IR)-active phonons. Since THz light encounters relatively low refractive index values ($n\approx2$) \cite{Ulian_2020_data} and an absence of free carriers in clinochlore, bulk samples can be effectively characterized using terahertz time-domain spectroscopy (THz-TDS) \cite{Koch_2023,Kawahala_2023} with minimal transmission losses. Additionally, a THz-TDS setup can be readily modified for polarization-sensitive experiments capable of discriminating anisotropic or chiral effects \cite{Norris_2019}. In this work, we performed THz polarimetry measurements to determine how bulk clinochlore reshapes the complex polarization state of transmitted THz waves. We observed an anisotropic, sharp and prominent resonance feature in the retrieved complex refractive index spectrum, which we identified as the manifestation of clinochlore rigid-layer shear phonon modes, predicted by theoretical computations. Its clear Lorentzian shape allowed the extraction of the effective phonon parameters as a function of sample orientation. Moreover, we show that the anisotropic modes introduce complex Faraday rotations near the resonance, through detected differences between the opposite handedness of the transmitted THz waves' circular polarization components. Finally, our analysis indicates that the observed anisotropic behavior of shear phonons can be accurately described using a Jones formalism approach, with a matrix of linear eigenstates \cite{Menzel_Rockstuhl_Lederer_2010}.

\section*{RESULTS}
\paragraph*{Polarization-sensitive time-domain waveforms}\mbox{}

The structure of clinochlore is a variation of talc \cite{Oliveira_2024}, a well-behaved phyllosilicate mineral that serves as a standard model for 2:1 trioctahedral phyllosilicates. Talc is composed of two tetrahedral silicon oxide layers with apical oxygens pointing towards a magnesium trioctahedral layer positioned between them, with oxygen (O) and hydroxyl (OH) groups at the vertices \cite{Deer_Howie_Zussman_2013}. In clinochlore, one of the four silicon ions in the talc-like layer is substituted by aluminum ions \cite{Moro_Ulian_Valdre_2016}, leading to charge imbalance. Charge neutrality is restored by the formation of an additional trioctahedral brucite-like layer that intercalates between the talc-like layers \cite{Deer_Howie_Zussman_2013}. This brucite-like layer has a mixed contribution of magnesium and aluminum as central ions \cite{Oliveira_2022}. Clinochlore exhibits a monoclinic crystal structure \cite{Gopal_2004}, as shown in Fig.~1A, with the intercalated brucite- and talc-like layers stacked along the $c$ axis, parallel to the $ab$ plane.

Our sample was previously characterized by other techniques \cite{Oliveira_2022}, which found that the most common impurity is iron in different oxidation states (Fe$^{2+}$ or Fe$^{3+}$). Considering this, we performed density functional perturbation theory (DFPT) of the lattice vibrations of bulk clinochlore, starting from the pristine structure Mg$_5$Al(AlSi$_3$)O$_{10}$(OH)$_8$, and further introducing iron ions into octahedral sites. Our DFPT results predicted two IR-active rigid-layer shear phonon modes with frequencies very close together, below 1.65~THz (55~cm$^{-1}$) for the pristine structure. Those frequencies are decreased towards 1.14~THz (38~cm$^{-1}$) when the iron impurities are introduced; see the Supplementary Materials~\cite{SM}. For these phonons, adjacent layers move rigidly in opposite directions with respect to one another, as Fig.~1B illustrates for the lower (red arrows along the $a$ axis) and the higher (blue arrows along the $b$ axis) frequency modes. 

To fully record the optical anisotropic behavior, we modified a conventional setup for THz-TDS in a transmission geometry \cite{Kawahala_2023} by adding wire grid polarizers (WGP) to enable polarization-sensitive measurements \cite{Morikawa_2006}, as presented in the scheme of Fig.~1C. For a fixed incident THz polarization direction, parallel to the lab $x$ axis, the transmitted THz electric field temporal waveforms were collected in an orthogonal basis of $E_x(t)$ and $E_y(t)$; see Fig.~1D. Because amplitude and phase components are preserved in THz-TDS measurements, the acquired waveforms carry information of both polarization rotation $\theta$ and ellipticity $\eta$ angles. As shown in Fig.~1D, although most of the transmission amplitude remains aligned with the incident THz polarization, a measurable non-negligible $y$ component introduces an overall polarization rotation and change in ellipticity. These measurements were repeated for different rotation angles $\beta$ of the sample crystal plane $ab$ around lab $z$ axis; see the Suplementary Materials~\cite{SM}. Further, spectral information was recovered from time-domain data by Fourier transforming it into frequency ($\nu$)-dependent complex-valued THz electric fields $E_x(\nu)$ and $E_y(\nu)$.

\paragraph*{Identification of the phonon mode}\mbox{}

Before dealing with the full polarization state, it is insightful to first examine the information carried by the $x$-polarized component of the THz-TDS measurements, which contains most of the signal amplitude, as noted above. For arbitrary $\beta$, Fig.~2A compares $E_x(t)$ collected for clinochlore and talc samples (blue and orange curves, respectively), along with a reference waveform $E_\textrm{ref}(t)$ measured in the absence of any sample (gray curve). The Fourier amplitudes $|E_x|(\nu)$ of each signal are shown in Fig.~2B, where a strong and sharp absorption pattern, centered around 1.13~THz, is observed exclusively for clinochlore. Conventional transmission models in the bulk limit \cite{Neu_Schmuttenmaer_2018} were then used to derive the complex refractive index spectrum $\tilde{n}_{xx}(\nu)=n_{xx}(\nu)+i\kappa_{xx}(\nu)$ of each material from their transmission coefficients $T_x(\nu)=E_x(\nu)/E_\textrm{ref}(\nu)$. Fig.~2C shows the frequency dependence of the real part ($n_{xx}$). From the imaginary part, namely the extinction coefficient $\kappa_{xx}$, we calculated the absorption coefficients $\alpha_{xx}=4\pi\nu\kappa_{xx}/c$, where $c$ is the speed of light in vacuum; see Fig.~2D. Clinochlore $\tilde{n}_{xx}$ spectrum exhibits a clear Lorentz resonance feature, typical of a phonon mode, which is absent in talc.

Given that low-frequency intralayer IR modes in clinochlore are expected to have very low intensities \cite{Ulian_2020}, the retrieved high values of $\alpha_{xx}$ disfavor hypotheses of in-layer mechanisms for the origin of the observed resonance. A comparison with the talc sample results also supports this statement, at least as it concerns the talc-like layer in clinochlore. Conversely, interlayer mechanisms are endorsed by the observed resonance central frequency, because it matches the range predicted by our DFT calculations of rigid-layer shear phonons in bulk clinochlore with iron impurities.

\paragraph*{Orientation-dependent effective phonons}\mbox{}

Now, for the clinochlore sample, we investigate the orientation dependence of $\tilde{n}_{xx}(\nu,\beta)$, by analyzing the $x$-polarized transmission data for different angles $\beta$ ranging from 0° to 360°. Here, we consider $\beta=0$° the orientation in which laboratory and sample coordinates are coincident. Restricting our analysis to the THz spectrum around the resonance, the experimental curves obtained for $n_{xx}(\nu,\beta)$ and $\kappa_{xx}(\nu,\beta)$ were used to compose the polar maps shown in Fig.~3A and Fig.~3B, respectively. In these visual representations, the frequency values increase radially, and the $\beta$ angles are arranged around the circle (azimuth).

Examining first the map of $n_{xx}(\nu,\beta)$ in Fig.~3A, we observe two main regions delimited by a white-colored oval-shaped contour that is flattened on the sides. The inner red-colored region (lower frequencies) indicates refractive index values higher than those in the outer blue-colored area (higher frequencies). Given the Lorentzian shape of the resonances, the white contour curve should be closely related to the orientation-dependent central frequencies of the effective modes $\nu_\textrm{eff}(\beta)$. Thus, our results suggest the detection of effective phonons with lower frequencies along the $a$ direction, increasing slightly as the orientation shifts towards the $b$ direction. Modeling with effective modes enables us to account for the averaging over grains with different crystallographic orientations in our sample. The anisotropic behavior of the refractive index is also consistent with our expectations for the shear modes.

A similar analysis can be drawn from the polar map of $\kappa_{xx}(\nu,\beta)$, presented in Fig.~3B. The red-colored doughnut-shaped pattern corresponds to the absorption spectrum near the peak, for different orientations of the clinochlore sample. Once again, the flattening on the sides indicates the anisotropic behavior of the effective phonon frequencies. Moreover, the narrow spectral range of the maps points out that the $\beta$-dependent variation of $\nu_\textrm{eff}$ is actually very small. In the following, we quantify it in terms of permittivity models.

The complex refractive index $\tilde{n}$ of a material is connected to its electric properties through the complex permittivity (dielectric function) $\epsilon=\tilde{n}^2$. In the absence of free carrier effects, we can model the permittivity spectrum as a damped Lorentz oscillator \cite{Fox_2010} to extract information from the $\beta$-dependent effective phonon modes as
\begin{equation}\label{eq:lorentz}
    \tilde{n}_{xx}^2(\nu,\beta) = \epsilon_{xx}(\nu,\beta) = \epsilon_\infty + \frac{(\epsilon_\textrm{s} - \epsilon_\infty)\nu_\textrm{eff}^2(\beta)}{\nu_\textrm{eff}^2(\beta)-\nu^2-i\nu\gamma_\textrm{eff}(\beta)},
\end{equation}
where $\epsilon_\infty$ and $\epsilon_\textrm{s}$ are the high-frequency and static dielectric constants, and $\nu_\textrm{eff}(\beta)$ and $\gamma_\textrm{eff}(\beta)$ are the orientation-dependent effective phonon central frequency and linewidth, respectively.

Let us consider a subset of the data presented in the polar maps examined above. The plots from Fig.~3C and Fig.~3D show, respectively, the experimental curves (hollow squares) of $n_{xx}(\nu)$ and $\kappa_{xx}(\nu)$ for selected values of $\beta$ between 0° and 100°. The curves were vertically offset for clarity. For each angle, the model from equation \ref{eq:lorentz} was used in the simultaneous fit of $n_{xx}$ and $\kappa_{xx}$, where all parameters were varied during the optimization process. The resulting optimal curves (black lines) in Fig.~3C and Fig.~3D show excellent agreement with the experimental data.

The optimization process was repeated for all probed $\beta$ angles. From the optimal parameters, the mean values of the dielectric constants were determined as $\epsilon_\infty=4.17\pm0.01$ and $\epsilon_\textrm{s}=4.28\pm0.01$, while the behaviors of $\nu_\textrm{eff}(\beta)$ and $\gamma_\textrm{eff}(\beta)$ (hollow squares) are presented in the respective plots of Fig.~3E and Fig.~3F. In these figures, the black line curves indicate that the experimental effective phonon parameters are well approximated by the models
\begin{equation}\label{eq:eff_pars}
\left\{\begin{aligned}
    \nu_\textrm{eff}^2(\beta) &= \nu_a^2\cos^2\beta + \nu_b^2\sin^2\beta,\\
    \gamma_\textrm{eff}^{-1}(\beta) &= \gamma_a^{-1}\cos^2\beta + \gamma_b^{-1}\sin^2\beta,
\end{aligned}\right.
\end{equation}
where we have defined $\nu_a\equiv\nu_\textrm{eff}(\beta=0\textrm{°})$, $\nu_b\equiv\nu_\textrm{eff}(\beta=90\textrm{°})$, and similarly for the linewidths~\cite{SM}. It then follows that the effective phonons can be described as a mixture of two modes: one aligned along the $a$ direction, with $\nu_a=(1.129\pm0.001)$~THz and $\gamma_a=(0.059\pm0.001)$~THz; and the other aligned along the $b$ direction, with a slightly higher frequency $\nu_b=(1.135\pm0.001)$~THz and a smaller linewidth $\gamma_b=(0.053\pm0.001)$~THz. 

\paragraph*{Anisotropy-induced complex Faraday rotations}\mbox{}

We analyzed the full polarization state of clinochlore THz-TDS data in a circularly polarized basis to track changes in the terahertz polarization axis and ellipticity. Therefore, we transformed the measured spectra $E_x(\nu,\beta), E_y(\nu,\beta)$ into left- (LHCP) and right-hand circular polarization (RHCP) components as $E_\ell(\nu,\beta)=\big(E_x(\nu,\beta)-iE_y(\nu,\beta)\big)/\sqrt{2}$ and $E_r(\nu,\beta)=\big(E_x(\nu,\beta)+iE_y(\nu,\beta)\big)/\sqrt{2}$, respectively \cite{Li_Yoshioka_2019}. With this approach, we can examine how clinochlore transmission reshapes both LHCP and RHCP components of the incident linearly polarized THz spectrum. For instance, if we define the transmission coefficients $T_\ell(\nu,\beta)=E_\ell(\nu,\beta)/E_\textrm{ref}(\nu)$ and $T_r(\nu,\beta)=E_r(\nu,\beta)/E_\textrm{ref}(\nu)$, we can quantify the transmission response differences for each handedness. Fig.~4A presents the behavior of $-2\log|T(\nu,\beta)|$ near the resonance center, for LHCP (solid lines) and RHCP (dashed lines) data, across $\beta$ angles from 0° to 180°. The observed peaks are related to the total absorption responses due to the shear modes.

When the incident THz polarization is closely aligned with one of the sample axes, $a$ or $b$ (for example, $\beta=$ 0°, 100°, or 180° in Fig.~4A), the differences in LHCP and RHCP absorptions are minimal. On the other hand, for sample orientations between these directions, we observe not only different amplitudes, but also frequency shifts when comparing the components. Given that these effects are relatively symmetrical, and considering the anisotropic behavior of the phonon modes discussed in the previous section, we can employ geometric reasoning to infer that the circular transmission coefficients may be modeled as
\begin{equation}\label{eq:circ_T}
\left\{\begin{aligned}
    T_\ell(\nu,\beta) &= \frac{1}{\sqrt{2}}\Big(T_a(\nu)\cos\beta+iT_b(\nu)\sin\beta\Big)e^{-i\beta},\\
    T_r(\nu,\beta) &= \frac{1}{\sqrt{2}}\Big(T_a(\nu)\cos\beta-iT_b(\nu)\sin\beta\Big)e^{i\beta},
\end{aligned}\right.
\end{equation}
where $T_a(\nu)$ and $T_b(\nu)$ are the linear transmission coefficients due to the shear phonons along $a$ (lower frequency) and $b$ (higher frequency) axes, respectively. 

By inputting experimental data for $T_a(\nu)$ and $T_b(\nu)$ into equation \ref{eq:circ_T}, we reconstructed the curves shown in Fig.~4B. As we can see, the absorptions calculated from the model preserve the key features of the experimental data (Fig.~4A). Thus, this suggests that the observed polarization modification effects near the resonance may be associated with phonon anisotropy. A previous study on a seraphinite gemstone \cite{Han_2015} also demonstrated anisotropic absorption in that mineral variant.

Another way of investigating the polarization effects is by determining the complex Faraday rotations (FR), which we denote as $\tilde{\theta}$. One of the advantages of this approach is that it does not rely on a reference measurement, as it compares directly the LHCP and RHCP components of the signal through $\tilde{\theta} = -\arctan[i(E_r-E_\ell)/(E_r+E_\ell)]$ \cite{Cheng_Taylor_Folkes_Rong_Armitage_2019}. Written as $\tilde{\theta}(\nu,\beta)=\theta(\nu,\beta)+i\eta(\nu,\beta)$, the real part of the FR is actually the polarization rotation angle, and its imaginary part the ellipticity angle \cite{Hernandez_Baydin_2023}. For the same $\beta$ values and spectral range shown in Fig.~4A and Fig.~4B, the experimental curves for $\theta(\nu,\beta)$ and $\eta(\nu,\beta)$ are presented in the plots from Fig.~4C and Fig.~4D, respectively. As we can see, there are significant effects on both rotation and ellipticity data, centered around 1.13~THz, only for $\beta$ angles in which LHCP and RHCP absorptions are different.

We can better visualize this pattern through the polar maps shown in Fig.~4E and Fig.~4F. For each respective map, higher values of rotation or ellipticity are represented by more intense red or blue colors, depending on the sign. On the other hand, white regions are associated to zero rotation or ellipticity. In particular, the maps show that virtually no polarization modification effects occur for any radial orientation that matches clinochlore $a$ or $b$ axes. Therefore this gives us a important insight into the symmetries of the observed shear phonons, as $a$ and $b$ should be the eigenvectors of polarization.

\section*{DISCUSSION}

We can use Jones calculus \cite{Fowles_1989} to understand the interaction of the THz light with materials in polarization-sensitive THz-TDS experiments. If we consider the polarization states of the incident and transmitted THz electric fields as the respective Jones vectors $\mathbf{E}_\textrm{i}(\nu)$ and $\mathbf{E}_\textrm{t}(\nu)$, they may be related through a frequency-dependent $2\times2$ Jones matrix $\mathbf{T}(\nu)$ so that $\mathbf{E}_\textrm{t}(\nu)=\mathbf{T}(\nu)\mathbf{E}_\textrm{i}(\nu)$ \cite{Armitage_2014}. The information gathered from our experimental results suggests that, considering the anisotropic shear phonons in clinochlore, the Jones transmission matrix may take the form
\begin{equation}\label{eq:T_ab}
    \mathbf{T}(\nu) = \begin{bmatrix}
        T_a(\nu) & 0\\
        0 & T_b(\nu)
    \end{bmatrix},
\end{equation}
when written in the basis of the eigenvectors of polarization, $a$ and $b$. Thus, the associated reduced ($2\times2$) dielectric tensor $\boldsymbol{\epsilon}(\nu)$ \cite{Vernon_Huggins_1980} is also diagonal in this basis
\begin{equation}\label{eq:eps_ab}
    \boldsymbol{\epsilon}(\nu) = \begin{bmatrix}
        \epsilon_a(\nu) & 0\\
        0 & \epsilon_b(\nu)
    \end{bmatrix},
\end{equation}
where, $\epsilon_a(\nu)$ and $\epsilon_b(\nu)$ are, respectively, the permittivity spectra of the lower and the higher frequency phonon modes. Moreover, we consider the non-zero elements of $\boldsymbol{\epsilon}(\nu)$ being described by independent Lorentz oscillators.

On the other hand, our THz-TDS experimental setup is mainly described in terms of the fixed laboratory coordinates $xy$. For instance, to recover the effective dielectric response along $x$ axis we can compute $\epsilon_{xx}=\hat{x}\cdot(\boldsymbol{\epsilon}\cdot\hat{x})$, where $\hat{x}$ is the unit vector in this direction. Naturally, $\epsilon_{xx}$ now depends on the sample orientation, so that
\begin{equation}\label{eq:eff_phonon}
    \epsilon_{xx}(\nu,\beta) = \epsilon_a(\nu)\cos^2\beta + \epsilon_b(\nu)\sin^2\beta.
\end{equation}
We then notice that, as long as the phonon central frequencies $\nu_a$ and $\nu_b$ are sufficiently close together (which is the case for our observations), $\epsilon_{xx}(\nu,\beta)$ itself may be approximated as a Lorentz oscillator, regardless of the $\beta$ value. In this sense, with the additional assumption for the linewidths $\gamma_a\approx\gamma_b$, equations \ref{eq:lorentz} and \ref{eq:eff_phonon} predict the phenomenological models for  $\nu_\textrm{eff}$ and $\gamma_\textrm{eff}$ introduced in equation \ref{eq:eff_pars}.

To express the Jones matrix in laboratory coordinates, we need to rotate it an angle $\beta$ using the transformation $\mathbf{T}(\nu,\beta)=\mathbf{R}^{-1}(\beta)\mathbf{T}(\nu)\mathbf{R}(\beta)$, where $\mathbf{R}(\beta)$ is the unitary matrix of rotations \cite{Menzel_Rockstuhl_Lederer_2010}, and the $\beta$ dependence of $\mathbf{T}(\nu,\beta)$ implies it is no longer in the basis of the eigenvectors. Hence, the rotated Jones matrix can be written as
\begin{equation}\label{eq:rot_T}
    \mathbf{T}(\nu,\beta)=\begin{bmatrix}
        T_a(\nu)\cos^2\beta+T_b(\nu)\sin^2\beta & \Big(T_a(\nu)-T_b(\nu)\Big)\sin\beta\cos\beta \\
        \Big(T_a(\nu)-T_b(\nu)\Big)\sin\beta\cos\beta & T_a(\nu)\sin^2\beta+T_b(\nu)\cos^2\beta
    \end{bmatrix},
\end{equation}
which has the generalized shape of a simple anisotropic media with linear eigenstates, indicating a mirror symmetry $M_{nz}$ with respect to a plane that contains $z$ axis \cite{Armitage_2014}. Further, we may compute $\mathbf{T}_\textrm{circ}(\nu,\beta)=\boldsymbol{\Lambda}^{-1}\mathbf{T}(\nu,\beta)\boldsymbol{\Lambda}$, where $\boldsymbol{\Lambda}$ is the unitary transformation matrix to a circular polarization basis \cite{Menzel_Rockstuhl_Lederer_2010}, resulting in
\begin{equation}\label{eq:jones_circ}
    \mathbf{T}_\textrm{circ}(\nu,\beta) = \frac{1}{2}\begin{bmatrix}
        T_a(\nu) + T_b(\nu) & \Big(T_a(\nu) - T_b(\nu)\Big)e^{-2i\beta}\\
        \Big(T_a(\nu) - T_b(\nu)\Big)e^{2i\beta} & T_a(\nu) + T_b(\nu)
    \end{bmatrix}.
\end{equation}

As we can see, given that the diagonal elements in equation \ref{eq:jones_circ} are equal, there is no chirality, but the anisotropy is encoded in the different off-diagonal elements. Additionally, the same transmission coefficients derived in equation~\ref{eq:circ_T} can be retrieved from $\mathbf{T}_\textrm{circ}$ through $T_\ell(\nu,\beta)=\hat{\ell}\cdot(\mathbf{T}_\textrm{circ}(\nu,\beta)\cdot\hat{x})$ and $T_r(\nu,\beta)=\hat{r}\cdot(\mathbf{T}_\textrm{circ}(\nu,\beta)\cdot\hat{x})$, where $\hat{\ell}$ and $\hat{r}$ are the respective unit vectors of LHCP and RHCP~\cite{SM}. Thus, our approach demonstrates optical anisotropy arising from material resonances, in contrast to the anisotropy reported for vdW materials in the visible range due to geometrical anisotropy of the elementary crystal cell \cite{Slavich_2024}, as well as for controlled arrangement of two-dimensional \cite{voronin_2024_2} and quasi-one-dimensional structures \cite{Voronin_2024}.

In conclusion, we studied the THz transmission spectrum of a bulk clinochlore sample, retrieved through polarization-sensitive THz-TDS measurements. From the experimental data, we determined the static and high-frequency dielectric constants with values of $4.28\pm0.01$ and $4.17\pm0.01$, respectively.  We observed a strong and sharp anisotropic absorption line, centered around 1.13~THz. Regardless of the clinochlore sample $ab$ plane orientation around $z$ axis, its effective complex refractive index spectrum exhibited great agreement with a damped Lorentz oscillator model, characteristic of a phonon mode. Comparisons with DFPT calculations were used to identify this resonance as the clinochlore rigid-layer shear phonon dynamics, with a mode frequency that can be tuned from the pristine case (1.65~THz) to our case when considering the incorporation of iron impurities. Analyses of the experimental data indicate that our observations can be explained in terms of effective phonon modes with a lower frequency (1.129~THz) linear mode aligned to the $a$ axis, and another one aligned to the $b$ axis with a higher frequency (1.135~THz). Even though these frequencies were found to be closer together than those from the theoretical predictions, the anisotropy of the modes was sufficient to induce Faraday rotation and ellipticity. Moreover, using Jones formalism the shear phonons contribution to clinochlore THz transmission spectrum was successfully encoded in a Jones matrix with linear eigenstates, like in a simple anisotropic media. Considering that we are working with a large polycrystalline mineral sample from a natural environment, the reported anisotropy in the terahertz range is remarkable. These results demonstrate that fundamental information about interlayer vibrational modes can be obtained from the terahertz optical anisotropy in van der Waals minerals. Applications of the observed shaping of the polarization of transmitted light pulses reinforces the potential of using abundant layered minerals as the source of two-dimensional materials with controllable phononic properties through impurity engineering.

\section*{MATERIALS \& METHODS}

\paragraph*{Samples}\mbox{}

For the main investigations in this work, we used a natural bulk clinochlore sample, extracted from Minas Gerais/Brazil geological environment, shaped as a 140~$\upmu$m-thick nearly square slab with side dimensions on the order of 3~mm. Given that phyllosilicates are hydrated minerals, the clinochlore sample surpassed a heat treatment in a tubular furnace under ambient atmosphere for 90 min at 500°C to remove water content \cite{Oliveira_2024} and improve the sample crystallinity prior to the THz-TDS measurements. For the comparison measurement pointed out in the text, we used a natural bulk talc sample from the same geological environment and similar shape and dimensions.

\paragraph*{Measurements}\mbox{}

We used a conventional THz-TDS setup in a transmission geometry \cite{Kawahala_2023}, modified with a set of three wire grid polarizers (WGP) \cite{Morikawa_2006} to enable the polarimetry measurements. Pulses from a mode-locked Ti:sapphire laser oscillator tuned to 780~nm, with a pulse duration of 130~fs and a repetition rate of 76~MHz, were used for pumping a biased photoconductive antenna (PCA), set for generating linearly-polarized THz radiation in $x$ direction. This polarization state was carefully ensured by a polarizer (WGP1). The THz beam was focused with a 3~mm spot diameter onto the sample positioned in a rotation mount. The transmitted beam leaving the back of the sample passed through a pair of polarizers (WGP2 and WGP3) before it was collected and probed by another optically-gated PCA. We calibrated WGP1 and WGP3 for the same fixed polarization direction. Then, we defined a THz-TDS measurement as comprised of two different scans: $E_{+}(t)$ with WGP2 set to $+$45° (with respect to WGP1/WGP3); and $E_{-}(t)$ with WGP2 set to $-$45°. Finally, the full polarization state in laboratory coordinates was recovered through $E_x(t)=E_{+}(t)+E_{-}(t)$ and $E_y(t)=E_{+}(t)-E_{-}(t)$.

\paragraph*{Density-functional perturbation theory calculations}\mbox{}

The spin-polarized density-functional theory (DFT)\cite{hohenberg1964inhomogeneous,kohn1965self-consistent} and density-functional perturbation theory (DFPT)\cite{baroni2001phonons} calculations were performed by using the \textsc{quantum-espresso} distribution \cite{giannozzi2009quantum,giannozzi2017advanced}. The calculations were performed by using the local exchange correlation (XC) functional with the Perdew-Burke-Ernzerhof's parametrization of the generalized gradient approximation (GGA) for solids (PBEsol) \cite{perdew2008restoring}. The valence electron-nucleus interactions were described by using DOJO \cite{setten2018pseudodojo} optimized norm-conserving Vanderbilt pseudopotentials (ONCVPSP) \cite{hamann2013optimized}, setting the kinetic energy cutoff in the wave-function (charge density) expansion to 46~$E_\textrm{h}$ (184~$E_\textrm{h}$). The geometries were relaxed until the forces on atoms and stress on the lattice were lower than 2.57~meV/\r{A} and 50~MPa, respectively, using Brillouin Zone sampling in a $6\times4\times2$ $\Gamma$-centered Monkhorst-Pack (MP) grid \cite{monkhorst1976special}. The electronic occupations were smoothed by using 2.5~m$E_\textrm{h}$ within the cold smearing method \cite{marzari1999thermal}.

\bibliography{ref}
\bibliographystyle{Science}

\noindent\textbf{Acknowledgments:} The authors thank Prof. M. A. Fonseca from the Federal University of Ouro Preto (UFOP) for supplying the clinochlore and talc minerals. R.d.O. acknowledges the sample preparation facilities at Federal University of Minas Gerais (UFMG). R.L. acknowledges computational time at CENAPAD-SP, CENAPAD-RJ and SDumont supercomputer. R.L. and J.R.-S. acknowledge the Instituto Nacional de Ciência e Tecnologia em Nanomateriais de Carbono (INCT Nanocarbono).

\noindent\textbf{Funding:} This work was supported by the São Paulo Research Foundation (FAPESP), Grants Nos. 2021/12470-8 and 2023/04245-0.  F.G.G.H. acknowledges financial support from Grant no. 306550/2023-7 of the National Council for Scientific and Technological Development (CNPq). N.M.K. acknowledges support from FAPESP Grant No. 2023/11158-6. I.D.B. acknowledges support from FAPESP Grant Nos. 2019/14017-9 and 2022/02901-4 and CNPq Grant No. 306170/2023-0. R.L. acknowledges support from CENAPAD-SP Grant No. 695, SDumont supercomputer Grant Nos. 223869 and 245220, the Minas Gerais State Research Support Foundation (FAPEMIG) Grant No. APQ-01553-22 and CNPq Grant No. 409300/2023-3. J.R.-S. acknowledges support from FAPEMIG Grant Nos. APQ-01922-21 and APQ-01638-23 and CNPq Grant Nos. 408319/2021-6 and 315650/2023-0. R.L. and J.R.-S. acknowledge support from FAPEMIG Grant Nos. APQ-03741-23, RED-00081-23, RED-00079-23 and RED-00223-23.

\noindent\textbf{Author Contributions:} N.M.K., I.D.B and F.G.G.H. conceived the project. R.d.O. prepared the samples. N.M.K. and D.A.M. built the polarization-sensitive terahertz spectrometer and performed the measurements under the supervision of F.G.G.H. R.L. performed and analyzed the density-functional perturbation theory calculations, with contributions from J.R.-S. N.M.K. analyzed the experimental data and prepared the manuscript both under the supervision of F.G.G.H. R.L. and J.R.-S. contributed to the analysis of the origin of low-frequency resonances. All authors discussed the results and commented on the manuscript. 

\noindent\textbf{Competing Interests:} The authors declare that they have no competing interests.

\noindent\textbf{Data and Materials Availability:} All data needed to evaluate the conclusions in the paper are present in the paper and/or the Supplementary Materials.

\clearpage
\begin{figure}[ht]
    \includegraphics{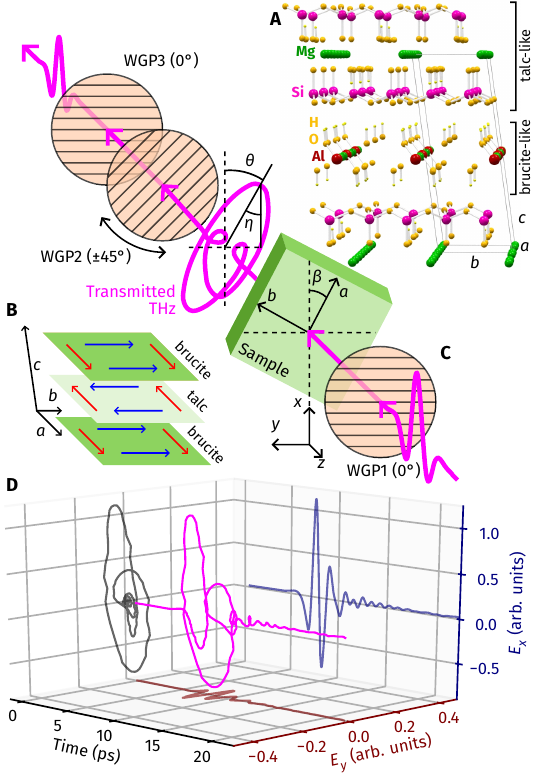}
    \caption{\textbf{Polarization-sensitive measurements.} (\textbf{A}) Representation of the crystal structure of clinochlore. (\textbf{B}) Illustration of the relative movement of adjacent brucite- and talc-like layers, in clinochlore, for the lower (red arrows) and the higher (blue arrows) frequency rigid-layer shear phonon modes. (\textbf{C}) Schematic diagram of the THz-TDS setup in transmission geometry with a configuration of wire grid polarizers (WGP) which allows polarization-sensitive measurements. (\textbf{D}) 
    Measured time-domain waveform showing the full polarization state of a THz pulse reshaped after being transmitted through the clinochlore sample.}
    \label{fig1}
\end{figure}

\clearpage
\begin{figure}[ht]
    \includegraphics{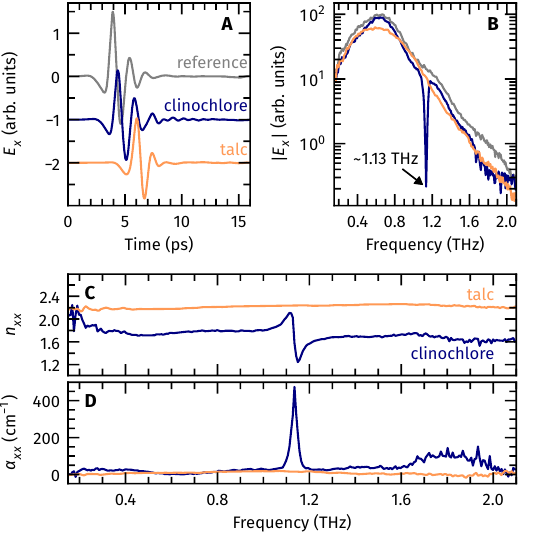}
    \caption{\textbf{Comparison of clinochlore and talc THz spectra.} (\textbf{A}) Time-domain waveforms of the THz electric field $x$-polarized component of pulses transmitted through clinochlore and talc samples, in comparison with a sample-free reference measurement. (\textbf{B}) Spectral amplitudes retrieved after Fourier-transforming the time-domain waveforms, indicating an absorption line in clinochlore THz spectrum. (\textbf{C}) Frequency-dependent refractive indices and (\textbf{D}) absorption coefficients extracted from THz-TDS experimental data for clinochlore and talc.}
    \label{fig2}
\end{figure}

\clearpage
\begin{figure}[ht]
    \makebox[\textwidth][c]{\includegraphics{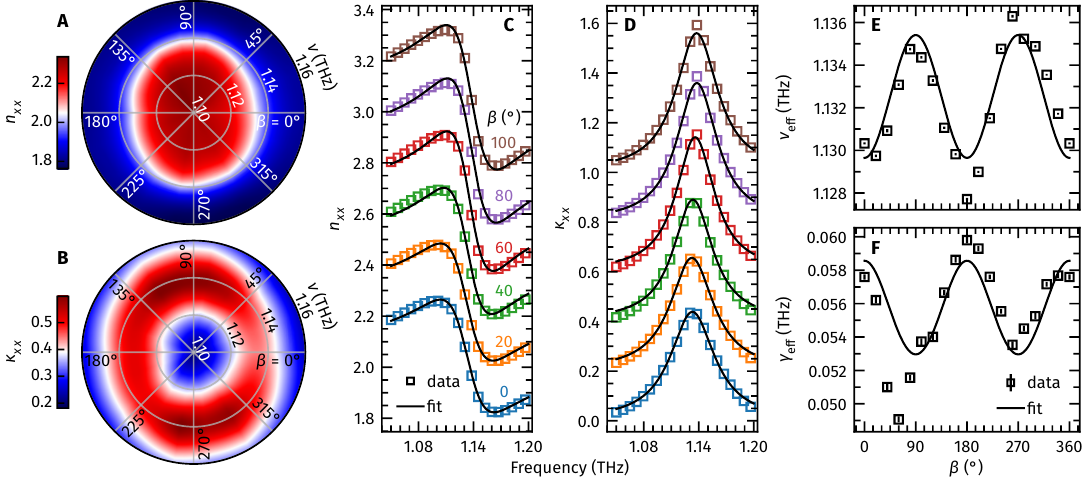}}
    \caption{\textbf{Effective phonons behavior.} Polar maps of (\textbf{A}) real and (\textbf{B}) imaginary parts of clinochlore effective refractive index spectrum, showing its orientation dependence (azimuth) for frequencies around the resonance (radial coordinate). (\textbf{C}) and (\textbf{D}) present the optimal fits of equation \ref{eq:lorentz} to the experimental data (hollow squares) for some angles $\beta$ in between 0° and 100°. (\textbf{E}) Optimized effective phonon central frequency and (\textbf{F}) linewidth parameters as a function of the orientation, showing the respective fits of equation \ref{eq:eff_pars}.}
    \label{fig3}
\end{figure}

\clearpage
\begin{figure}[ht]
    \makebox[\textwidth][c]{\includegraphics{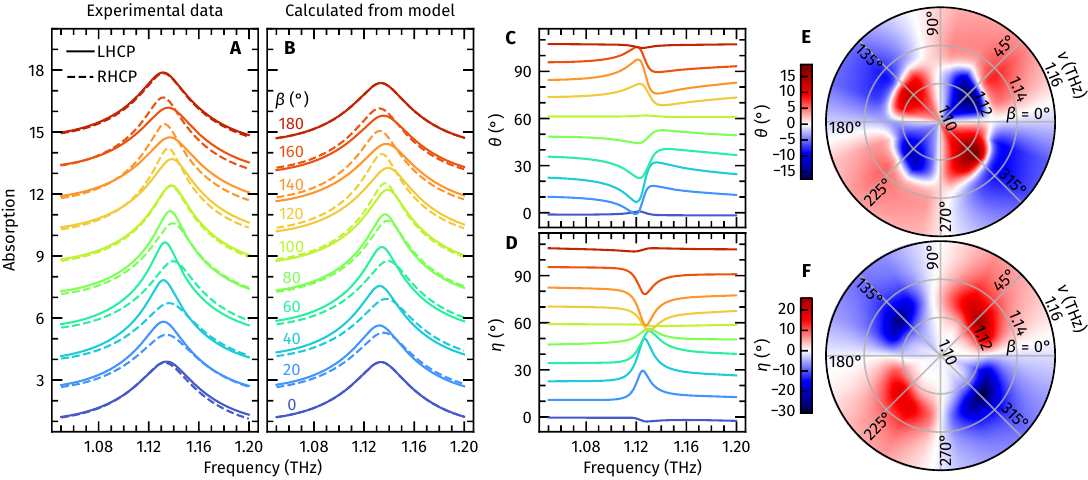}}
    \caption{\textbf{Orientation-dependent complex Faraday rotations.} (\textbf{A}) Comparison of $-2\log|T|$ determined from the LHCP (continuous lines) and RHCP (dashed lines) components of clinochlore THz transmission data, in the spectral range around the resonance, for some angles $\beta$ in between 0° and 180°. (\textbf{B}) The same quantities now calculated from model (equation \ref{eq:circ_T}). (\textbf{C}) real and (\textbf{D}) imaginary parts of the complex Faraday rotation. (\textbf{E}) and (\textbf{F}) present their respective polar map representations.}
    \label{fig4}
\end{figure}

\clearpage


\renewcommand{\thesection}{S\arabic{section}}   
\renewcommand{\thetable}{S\arabic{table}}   
\renewcommand{\thefigure}{S\arabic{figure}}
\renewcommand{\theequation}{S\arabic{equation}}

\renewcommand{\figurename}{Fig.}
\renewcommand{\tablename}{Table}

\setcounter{equation}{0}
\setcounter{figure}{0}
\section*{Supplementary Material}

In this Supplementary Material, we present details of the density-functional perturbation theory results, of the polarization-sensitive terahertz time-domain spectroscopy, optical coefficients extraction from THz-TDS data, computations in Jones formalism, fits of the effective permittivity, full spectral band results, and the temperature dependence of the effective modes.

\section{Details of the density-functional perturbation theory results}

In order to understand the cause of the lowest resonances ($\sim$1~THz) observed in the experiments, we simulated the rigid-layer modes of bulk clinochlore models without (i.e. pristine) and with iron impurities [iron(II) and iron(III) defects in the brucite-like layer, considering isoelectronic substitutions]. For the pristine case, we found $\nu_a=1.58$~THz and $\nu_b=1.62$~THz. The iron shifts the wavenumbers of the lowest resonances to lower values [$\nu_a=1.17$~THz and $\nu_b=1.34$~THz for iron(II); $\nu_a=1.14$~THz and $\nu_b=1.37$~THz for iron(III)]. These results can be rationalized by considering the wavenumbers given by $\omega=\sqrt{K/\mu}$, where $K$ accounts for interatomic force constant effects and $\mu$ for effective mass effects of the related atomic vibration pattern. In a first approximation, we neglect the $K$ (which may be of paramount importance  ({\it17}) and consider only the $\mu$. The results agree with the expectations that Fe\textsubscript{Al} or Fe\textsubscript{Mg} substitutions increase the $\mu$ and consequently decrease $\omega$. The wavenumbers of structures with iron content are in close agreement with the resonances found in our experiments. Thus, we attribute these resonances to the optical activity of the low-frequency rigid-layer shear modes in clinochlore, and that our samples contains iron impurities. 

\section{Details of the polarization-sensitive terahertz time-domain spectroscopy}
We used a terahertz time-domain spectroscopy (THz-TDS) setup for measurements in a transmission geometry. Both pulsed THz emission and detection were carried by photoconductive antennas (PCAs: Menlo Systems, and BATOP GmbH, respectively) optically pumped/gated by laser pulses from a mode-locked Ti:sapphire oscillator (76~MHz, 130~fs, 780~nm, Coherent Corp.). We used off-axis parabolic mirrors to collimate the divergent THz beam coming out of the emission PCA, as well as to focus it onto the sample with a 3~mm spot diameter, and to further collect and aim the transmitted beam towards the detection PCA. The polarization sensitiveness of the THz-TDS measurements was enabled by a set of three THz polarizers (Tydex, LCC). The first one ensured a horizontal polarization state $\ket{H}$ of the incident THz beam right before the sample (to correct eventual parabolic mirror-induced modifications of the polarization state of the emitted THz beam). We placed the second polarizer, which we denote \emph{analyzer}, right after the sample to select either the diagonal $\ket{+}\equiv(\ket{H}+\ket{V})/\sqrt{2}$ or the anti-diagonal $\ket{-}\equiv(\ket{H}-\ket{V})/\sqrt{2}$ components of the transmitted beam polarization state $E=E_H\ket{H}+E_V\ket{V}$ (where $\ket{V}$ is the vertical polarization state), so that $E_\pm=\ket{\pm}\bra{\pm}E=(1/\sqrt{2})(E_H\pm E_V)\ket{\pm}=(1/2)(E_H\pm E_V)(\ket{H}\pm\ket{V})$. Finally, the third polarizer selects the horizontal component of $E_\pm$ prior detection, in a way that we recover $E_H=\bra{H}E_{+}+\bra{H}E_{-}$ and $E_V=\bra{H}E_{+}-\bra{H}E_{-}$. Using the laboratory coordinate system described in the main text, we identify $E_H\to E_x$ and $E_V\to E_y$. When collecting the time-domain waveforms, we used a time step of 0.033~ps, with a time window of 10~ps (limited by characteristic back-reflections from THz optics), and a signal integration time of 30~ms. We performed fast Fourier transforms (FFT) of zero-padded temporal data to retrieve the complex THz spectra.

\subsection{Control of sample's orientation}
\begin{figure}[htb]
    \makebox[\textwidth][c]{\includegraphics{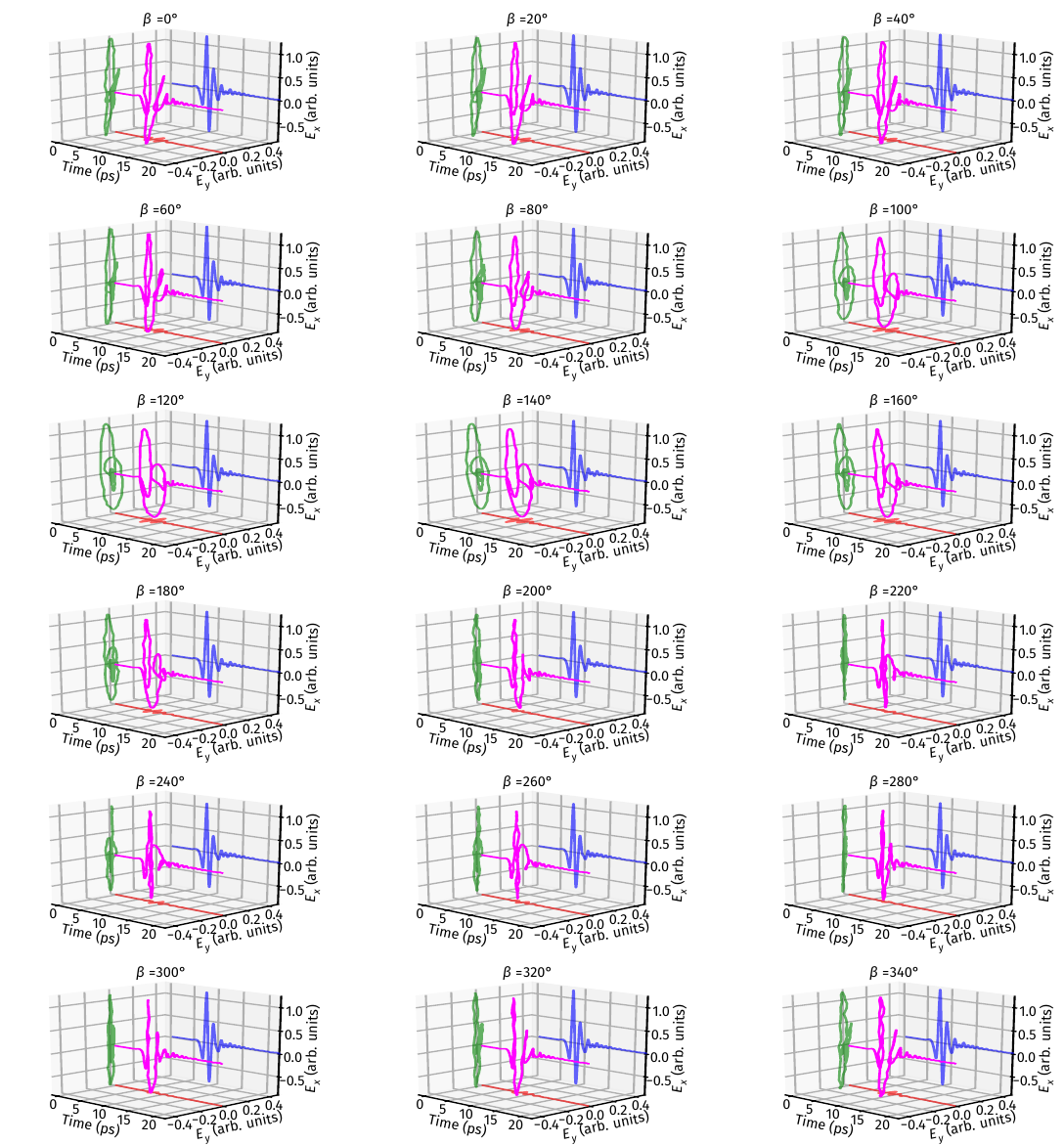}}
    \caption{\textbf{Orientation-dependent THz-TDS measurements output.} Full polarization states of the time-domain waveforms retrieved through orientation-dependent THz-TDS measurements of clinochlore transmission response.}
    \label{fig:smfig1}
\end{figure}

The clinochlore sample was placed in a rotation mount to allow the control of the azimuthal orientation of its $ab$ plane around $z$ axis. Considering the $\beta$ angles as defined in the main text, we performed THz-TDS measurements to retrieve the transmission full polarization state ($E_x, E_y$) for $\beta$ values in between 0° and 360°, with steps of 20°. A summary of the gathered data can be seen in the plots presented in Fig.~\ref{fig:smfig1}.

\section{Optical coefficients extraction from THz-TDS data}
If we consider homogeneous isotropic samples, their optical/electrical transmission response to THz electric fields is typically encompassed in a frequency ($\nu$)-dependent scalar quantity called transmission coefficient, defined by the ratio between sample and reference complex transmission spectra $T(\nu)=E_\textrm{sam}(\nu)/E_\textrm{ref}(\nu)$. The complex refractive index spectrum $\tilde{n}(\nu)=n(\nu)+i\kappa(\nu)$ of a bulk substrate-free sample ({\it32}) can be recovered in a good approximation through $n(\nu)=1+\arg[T]c/(2\pi\nu d)$ and $\kappa(\nu)=-\log[T(n+1)^2/4n)]c/(2\pi\nu d)$, where $c$ is the speed of light in vacuum, and $d$ is the sample thickness. The dielectric function of the sample (complex permittivity spectrum) is further determined through $\epsilon(\nu)=\tilde{n}^2(\nu)$. On the other hand, the transmission response of homogeneous anisotropic samples may be described by a Jones matrix $\mathbf{T}(\nu)$, so that
\begin{equation}\label{eqSM:T_matrix}
    \mathbf{E}_\textrm{sam} = \mathbf{T}\,\mathbf{E}_\textrm{ref} \qquad \Rightarrow \qquad \begin{bmatrix}
        E_\textrm{sam}^x \\ E_\textrm{sam}^y
    \end{bmatrix} = \begin{bmatrix}
        T_{xx} & T_{xy} \\ T_{yx} & T_{yy}
    \end{bmatrix} \begin{bmatrix}
        E_\textrm{ref}^x \\ E_\textrm{ref}^y
    \end{bmatrix},
\end{equation}
where each element $T_{lm}$ accounts for the $l$-polarized transmission response to an $m$-polarized source. Accordingly, the dielectric properties of the material may now be described by the permittivity tensor $\boldsymbol{\epsilon}(\nu)$.

\section{Details of the computations in Jones formalism}
In the main text, we discuss that the anisotropic shear phonons in clinochlore may be described by a diagonal Jones matrix when in crystal $ab$ basis; see equation~4. We rotate this matrix to laboratory $xy$ coordinates through the unitary transformation
\begin{equation}
\begin{aligned}
    \mathbf{T}(\beta) &= \mathbf{R}^{-1}(\beta)\,\mathbf{T}\,\mathbf{R}(\beta) = \begin{bmatrix}
        \cos\beta & -\sin\beta \\ \sin\beta & \cos\beta
    \end{bmatrix} \begin{bmatrix}
        T_a & 0 \\ 0 & T_b
    \end{bmatrix} \begin{bmatrix}
        \cos\beta & \sin\beta \\ -\sin\beta & \cos\beta
    \end{bmatrix} \\
    &= \begin{bmatrix}
            T_a\cos^2\beta+T_b\sin^2\beta & (T_a-T_b)\sin\beta\cos\beta \\
            (T_a-T_b)\sin\beta\cos\beta & T_a\sin^2\beta + T_b\cos^2\beta
    \end{bmatrix},
\end{aligned}
\end{equation}
i.e., the orientation ($\beta$)-dependent transmission coefficients in Cartesian coordinates are given by
\begin{equation}
    T_{xx} = T_a\cos^2\beta+T_b\sin^2\beta, \quad T_{yy} = T_a\sin^2\beta + T_b\cos^2\beta, \quad \textrm{and} \quad T_{xy}=T_{yx} = (T_a-T_b)\sin\beta\cos\beta.
\end{equation}
We further convert this matrix to the circular basis through the unitary transformation
\begin{equation}
\begin{aligned}
    \mathbf{T}_\textrm{circ}(\beta) &= \boldsymbol{\Lambda}^{-1}\,\mathbf{T}(\beta)\,\boldsymbol{\Lambda} = \frac{1}{\sqrt{2}}\begin{bmatrix}
        1 & -i \\ 1 & i
    \end{bmatrix} \begin{bmatrix}
        T_a\cos^2\beta+T_b\sin^2\beta & (T_a-T_b)\sin\beta\cos\beta \\
        (T_a-T_b)\sin\beta\cos\beta & T_a\sin^2\beta + T_b\cos^2\beta
    \end{bmatrix} \frac{1}{\sqrt{2}}\begin{bmatrix}
        1 & 1 \\ i & -i
    \end{bmatrix} \\
    &= \frac{1}{2}\begin{bmatrix}
        T_a + T_b & (T_a - T_b)e^{-2i\beta} \\ (T_a - T_b)e^{2i\beta} & T_a + T_b
    \end{bmatrix}_\textrm{circ} \equiv \begin{bmatrix}
        T_{\ell\ell} & T_{\ell r} \\ T_{r\ell} & T_{rr}
    \end{bmatrix}_\textrm{circ}.
    \end{aligned}
\end{equation}
We notice that the Jones transmission matrices (irrespectively of its representation, as it should be) indicates the presence of anisotropy, but no chirality. Moreover, in this circular basis, we write the reference electric field as $\mathbf{E}_\textrm{ref}=[E_\textrm{ref}, E_\textrm{ref}]_\textrm{circ}/\sqrt{2}$. Hence, our experiment may be described as
\begin{equation}
    \begin{bmatrix}
        E_\textrm{sam}^\ell \\ E_\textrm{sam}^r
    \end{bmatrix}_\textrm{circ} = \begin{bmatrix}
        T_{\ell\ell} & T_{\ell r} \\ T_{r\ell} & T_{rr}
    \end{bmatrix}_\textrm{circ} \frac{1}{\sqrt{2}}\begin{bmatrix}
        E_\textrm{ref} \\ E_\textrm{ref}
    \end{bmatrix}_\textrm{circ} = \frac{E_\textrm{ref}}{\sqrt{2}} \begin{bmatrix}
        T_{\ell\ell} + T_{\ell r} \\ T_{rr} + T_{r\ell}
    \end{bmatrix}_\textrm{circ},
\end{equation}
leading us to define the following quantities
\begin{equation}
    \left\{\begin{aligned}
        T_\ell &\equiv \frac{E_\textrm{sam}^\ell}{E_\textrm{ref}} = \frac{1}{\sqrt{2}}(T_{\ell\ell}+T_{\ell r}) = \frac{1}{\sqrt{2}}(T_a\cos\beta+iT_b\sin\beta)e^{-i\beta}\\
        T_r &\equiv \frac{E_\textrm{sam}^r}{E_\textrm{ref}} = \frac{1}{\sqrt{2}}(T_{rr}+T_{r\ell}) = \frac{1}{\sqrt{2}}(T_a\cos\beta-iT_b\sin\beta)e^{i\beta},
    \end{aligned}\right.
\end{equation}
which could also be obtained (less directly) through geometric reasoning, as pointed out in the main text, by working with projections of the THz electric fields throughout the different basis.

\section{Fits of the effective permittivity}
From the experimental analysis, we notice that the observed anisotropic behavior in the spectral region near the resonance can be modeled in terms of two orthogonal damped Lorentz oscillators, with central frequencies very close together, described by the dielectric functions
\begin{equation}
    \epsilon_a(\nu) = \epsilon_\infty^a + \frac{(\epsilon_\textrm{s}^a - \epsilon_\infty^a)\nu_a^2}{\nu_a^2-\nu^2-i\nu\gamma_a}, \qquad \textrm{and} \qquad \epsilon_b(\nu) = \epsilon_\infty^b + \frac{(\epsilon_\textrm{s}^b - \epsilon_\infty^b)\nu_b^2}{\nu_b^2-\nu^2-i\nu\gamma_b}.
\end{equation}
We further assume $\epsilon_\infty=\epsilon_\infty^a=\epsilon_\infty^b$ and $\epsilon_\textrm{s}=\epsilon_\textrm{s}^a=\epsilon_\textrm{s}^b$. As discussed in the main text, the effective permittivity we retrieve from $xx$ measurements is modeled as $\epsilon_{xx}(\nu,\beta)=\epsilon_a(\nu)\cos^2\beta+\epsilon_b(\nu)\sin^2\beta$, leading to
\begin{equation}\label{eqSM:exx_1}
    \epsilon_{xx}(\nu,\beta) = \epsilon_\infty + (\epsilon_\textrm{s}-\epsilon_\infty)\left[ \frac{\nu_a^2\cos^2\beta}{\nu_a^2-\nu^2-i\nu\gamma_a} + \frac{\nu_b^2\sin^2\beta}{\nu_b^2-\nu^2-i\nu\gamma_b} \right].
\end{equation}
Additionally, we saw that the frequencies are sufficiently close together for the effective permittivity itself behave as a damped Lorentz oscillator $\epsilon_{xx}=\epsilon_\infty+(\epsilon_\textrm{s}-\epsilon_\infty)\nu_\textrm{eff}^2/(\nu_\textrm{eff}^2-\nu^2-i\nu\gamma_\textrm{eff})$; see equation~1. We used this model to fit the experimental effective refractive index $\tilde{n}_{xx}(\nu,\beta)=\sqrt{\epsilon_{xx}(\nu,\beta)}$ data, recovering the optimized parameters $\epsilon_\infty$, $\epsilon_\textrm{s}$, $\nu_\textrm{eff}(\beta)$, and $\gamma_\textrm{eff}(\beta)$.

Moreover, for $\nu=\nu_\textrm{eff}$ equation~\ref{eqSM:exx_1} can be worked out as
\begin{equation}
\begin{aligned}
    \epsilon_{xx}(\nu_\textrm{eff},\beta) &= \epsilon_\infty + (\epsilon_\textrm{s}-\epsilon_\infty)\left[ \frac{\nu_a^2\cos^2\beta}{\nu_a^2-\nu_\textrm{eff}^2-i\nu_\textrm{eff}\gamma_a} + \frac{\nu_b^2\sin^2\beta}{\nu_b^2-\nu_\textrm{eff}^2-i\nu_\textrm{eff}\gamma_b} \right] \\ 
    &= \epsilon_\infty + (\epsilon_\textrm{s}-\epsilon_\infty)\frac{\nu_\textrm{eff}^2}{\nu_\textrm{eff}^2-\nu_\textrm{eff}^2-i\nu_\textrm{eff}\gamma_\textrm{eff}},
\end{aligned}
\end{equation}
which can be rearranged to the following approximation
\begin{equation}\label{eqSM:eff_qty}
    \frac{\nu_\textrm{eff}^2}{\gamma_\textrm{eff}} = \frac{\nu_a^2}{\gamma_a}\cos^2\beta + \frac{\nu_b^2}{\gamma_b}\sin^2\beta,
\end{equation}
considering that $\nu_a^2-\nu_\textrm{eff}^2\ll1$ and $\nu_b^2-\nu_\textrm{eff}^2\ll1$. Further approximations of equation~\ref{eqSM:eff_qty} lead to the models introduced in the main text (see equation~2), used for the fittings of $\nu_\textrm{eff}(\beta)$ and $\gamma_\textrm{eff}(\beta)$, which returned the optimized values of $\nu_a$, $\nu_b$, $\gamma_a$, and $\gamma_b$.

\section{Full spectral band results}
\begin{figure}[htb]
    \makebox[\textwidth][c]{\includegraphics{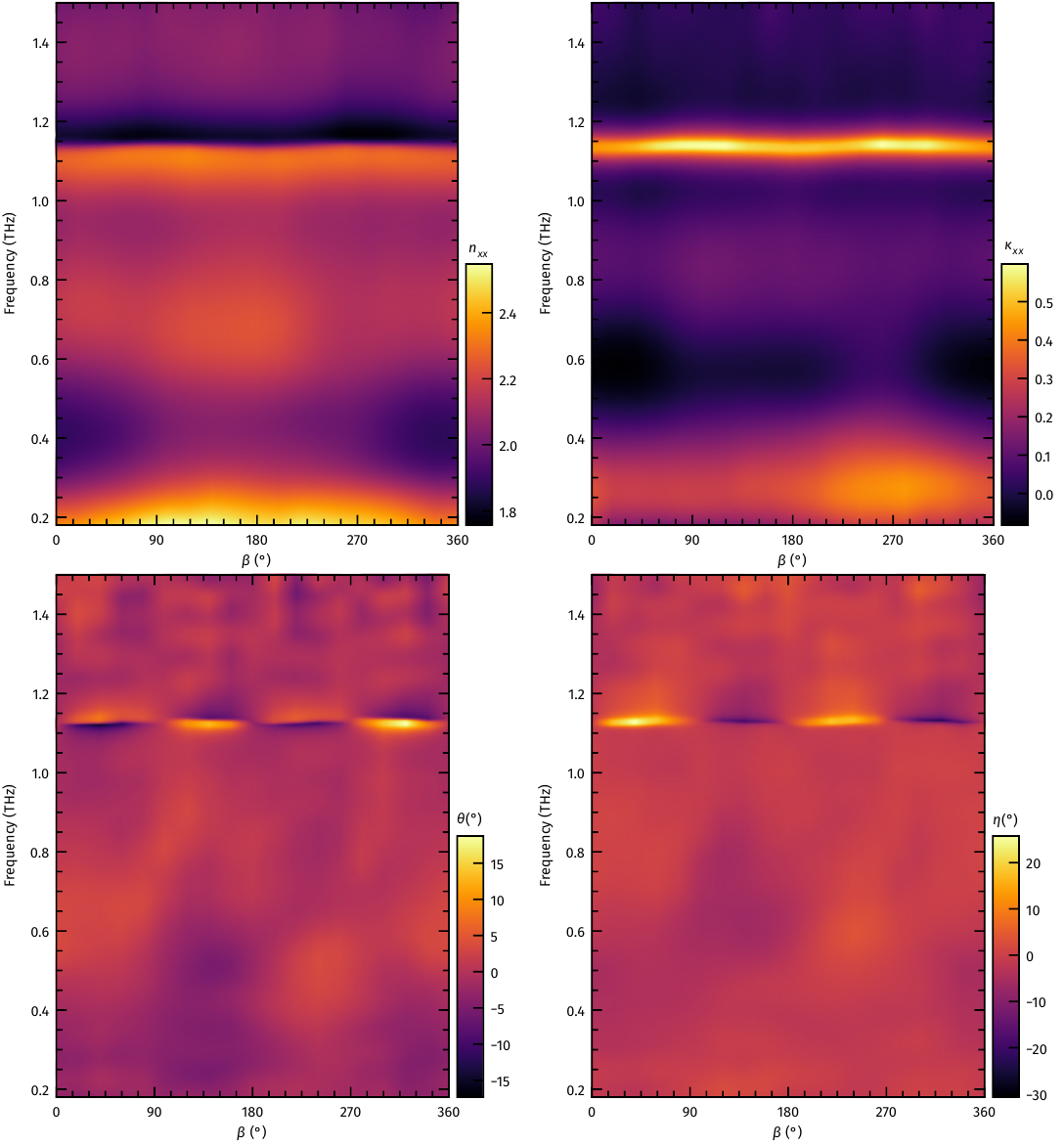}}
    \caption{\textbf{Experimental results in the full spectral band.} For frequencies in between 0.2~THz and 1.5~THz, the orientation ($\beta$)-dependent maps of the: (top left) effective refractive index $n_{xx}$; (top right) effective extinction coefficient $\kappa_{xx}$; (bottom left) Faraday rotation $\theta$; and (bottom right) Faraday ellipticity $\eta$.}
    \label{fig:smfig2}
\end{figure}

In the main text, we kept our analysis focused on the spectral region near the resonance, as studying the shear phonon modes is our main objective in this work. For completeness, Fig.~\ref{fig:smfig2} shows the maps for the complex refractive index $\tilde{n}_{xx}(\nu,\beta)=n_{xx}(\nu,\beta)+i\kappa_{xx}(\nu,\beta)$ and complex Faraday rotations $\tilde{\theta}(\nu,\beta)=\theta(\nu,\beta)+i\eta(\nu,\beta)$ in the full spectral band covered by our THz-TDS measurements (approximately in between 0.2~THz and 1.5~THz). Given the scarce low-frequency (THz band) optical characterization of clinochlore crystals found in literature, these maps can contribute with such information. For instance, below the resonance frequency, $n$ ranges from around 2.0 to 2.5, and $\kappa$ varies in between 0.0 and 0.4. On the other hand, above the resonance surroundings (up to 1.5~THz), there is virtually no absorption, while the refractive index is almost constant in 2.0. Furthermore, outside the resonance, we have not observed significant effects of Faraday rotation and ellipticity.

\section{Temperature dependence of the effective mode}
\begin{figure}[htb]
    \makebox[\textwidth][c]{\includegraphics{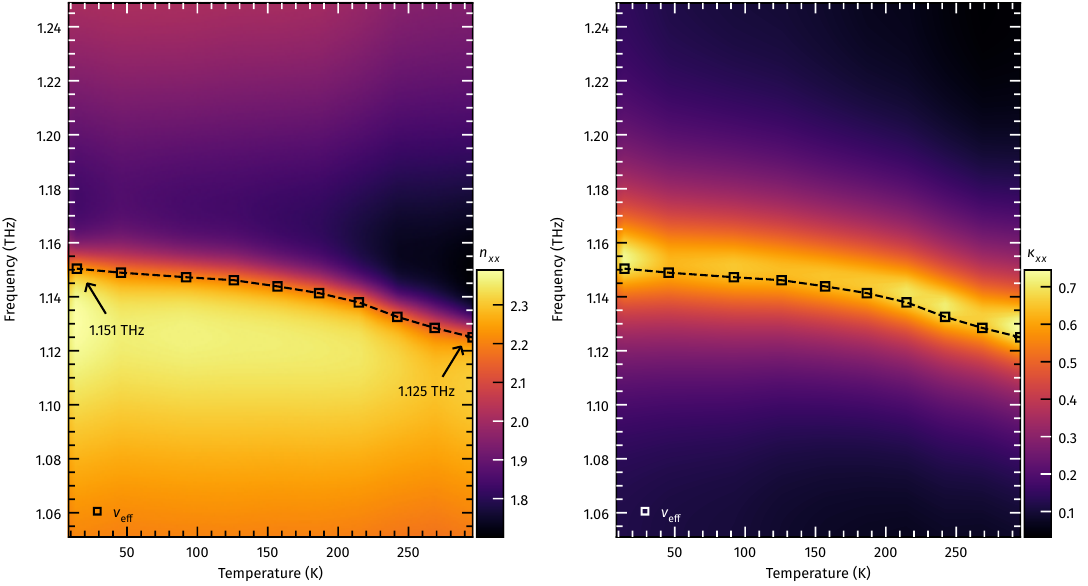}}
    \caption{\textbf{Temperature dependence of the effective mode.} For frequencies near the resonance, and a fixed orientation of the clinochlore sample, the temperature-dependent maps of the: (left) effective refractive index $n_{xx}$; and (right) effective extinction coefficient $\kappa_{xx}$. The hollow squares represent the fitted mode central frequencies $\nu_\textrm{eff}$ accompanied by a trend curve (dashed lines).}
    \label{fig:smfig3}
\end{figure}

In this section, we report temperature-dependent THz-TDS measurements of clinochlore, for a fixed sample orientation, in order to understand how the temperature influences the effective mode frequencies. For that, we mounted the clinochlore sample inside a cold finger cryocooler equipped with polytetrafluoroethylene windows. We retrieved the $x$-polarized component of time-domain waveforms at temperatures between 10~K to 300~K. For each temperature, we derived the complex refractive index $\tilde{n}_{xx}(\nu)$ spectrum, similarly as described in the main text. Fig.~\ref{fig:smfig3} shows the maps of the real (left) and imaginary (right) parts of $\tilde{n}_{xx}$ as a function of both temperature and frequency, near the resonance. Additionally, the hollow squares in the plots represent the effective mode central frequencies, retrieved through fits of the damped Lorentz oscillator model (equation~1 in the main text), for selected temperatures. The dashed lines are trend curves. We observe that $\nu_\textrm{eff}$ shows an increase of around 2~\% in its value as the temperature is decreased from 300~K ($\nu_\textrm{eff}^\textrm{300~K}=1.125$~THz) to 10~K ($\nu_\textrm{eff}^\textrm{10~K}=1.151$~THz), reflecting the increase in the interlayer forces at low temperature.

\clearpage

\end{document}